\documentclass[aps,twocolumn,pra,superscriptaddress,showpacs,tightenlines]{revtex4}
\usepackage{amssymb}
\usepackage{amsmath}
\usepackage{graphicx}
\usepackage{epsfig}
\usepackage{txfonts}
\usepackage{subfigure}
\usepackage{amsfonts}
\usepackage{CJK}

\begin{document}

\begin{CJK*}{GBK}{song}
\title{Impurity-induced Dicke quantum phase transition in an impurity-doped\\ cavity-Bose-Einstein condensate}
\author{Ji-Bing Yuan, Ya-Ju Song, and Le-Man Kuang\footnote{Author to whom any correspondence should be
addressed. Email: lmkuang@hunnu.edu.cn}}

\affiliation{Key Laboratory of Low-Dimensional Quantum Structures
and Quantum Control of Ministry of Education,  and Department of
Physics, Hunan Normal University, Changsha 410081, China}
\date{\today}

\begin{abstract}
We present a new generalized Dicke model, an impurity-doped Dicke
model (IDDM), by the use of an impurity-doped cavity-Bose-Einstein
condensate. It is shown that the impurity atom can induce Dicke
quantum phase transition (QPT) from the normal phase to superradiant
phase at a critic value of the impurity population. It is found that
the IDDM exhibits continuous Dicke QPT with an infinite number of
critical points, which is significantly different from that observed
in the standard Dicke model with only one critical point. It is
revealed that  the impurity-induced Dicke QPT can happen  in an
arbitrary coupling regime of the cavity field and atoms while the
Dicke QPT in the standard Dicke model occurs only in the strong
coupling regime of the cavity field and atoms. This opens a way to
observe the Dicke QPT in the intermediate and even weak coupling
regime of the cavity field and atoms.

\end{abstract}

\maketitle \narrowtext
\end{CJK*}

\section{\label{Sec:1}Introduction}

In recent years ultracold atoms in optical cavities have revealed
themselves as attractive new systems for studying
strongly-interacting quantum many-body theories. Their high degree
of tunability makes them especially attractive for this purpose. One
example, which has been extensively studied theoretically and
experimentally, is the Dicke quantum phase transition (QPT) from the
normal phase to the superradiant phase with a Bose-Einstein
condensate (BEC) in an optical cavity
\cite{Dimer,Baumann1,Keeling,Nagy1,Nagy2,Baumann2,Bastidas,Bhaseen,Liu1,Chen,Larson,Zhang,Liu2,Yuan,Konya}.
The Dicke model \cite{Dicke1954} describes a large number of
two-level atoms interacting with a single cavity field mode, and
predicts the existence of the Dicke QPT \cite{hepp,Sachdev1999} from
the normal phase to the superradiant phase. However, it is very hard
to observe the Dicke QPT in the standard Dicke model, since the
critical collective atom-field coupling strength needs to be of the
same order as the energy separation between the two atomic levels.
Fortunately, strong collective atom-field coupling has realized
experimentally in a BEC coupling with a ultrahigh-finesse cavity
filed \cite{Brennecke2007,Colombe2007}. Employing the cavity-BEC
system, the Dicke QPT  has been observed experimentally through an
atom-field coupling between a motional degree of freedom of the BEC
and the cavity field \cite{Baumann1}. The Dicke QPT corresponds to
the process of self-organization of atoms \cite{Nagy2008}. In the
experimental realization of the Dicke QPT based on the cavity-BEC
system \cite{Baumann1}, the normal phase corresponds to the BEC
being in the ground state associated with vacuum cavity field state
while  both the BEC and cavity field have collective excitations in
the super-radiant phase. A few extended Dicke models
\cite{Liu1,Chen,Zhang,Li,Chenqh} have been proposed to reveal rich
phase diagrams and exotic QPTs, which are different from those in
the original Dicke model.

In this paper, motivated by the recent experimental progress of
cavity-BEC and impurity-doped BEC systems \cite{Chikkatur,Zipkes} we
propose a generalized Dicke model, an impurity-doped Dicke model
(IDDM), by the use of an impurity-doped cavity-BEC. In our model,
the impurity atom is an internal two-level system. The impurity-BEC
interaction is tunable by an external magnetic field in the vicinity
of Feshbach resonances \cite{Ferlaino,Klempt}. The cavity-BEC system
adopted in our scheme is the same as that in the  Dicke QPT
experiment \cite{Baumann1}. The IDDM can reduce to the original
Dicke model when the impurity-BEC interaction is switched off. We
discuss how the presence of an impurity atom modifies the results of
the original Dicke model. We show that the impurity atom can induce
the Dicke QPT from the normal phase to the superradiant phase with
the impurity population being the QPT parameter. It is found that
the IDDM exhibits continuous Dicke QPT with an infinite number of
critical points, which is different from that observed in the
standard Dicke model with only one critical point. It is predicted
that the impurity-induced Dicke QPT can happen  in an arbitrary
coupling regime of the cavity field and atoms while the Dicke QPT in
the standard Dicke model occurs only in the strong coupling regime
of the cavity field and atoms. This opens a possibility to observe
the Dicke QPT in the intermediate and even weak coupling regime of
the cavity field and atoms. We propose a scheme to control the
impurity population in the cavity-BEC through making quantum
measurements on an auxiliary atom outside the cavity, which is
correlated to the impurity atom in the cavity-BEC.

The rest of the paper is organized as follows. In Sec. II, we
present the IDDM by the use of an impurity-doped cavity-BEC system.
In Sec. III, we investigate QPT properties of the IDDM and show the
presence of continuous  Dicke QPT with an infinite number of
critical points in the IDDM. We also analyze the impurity-population
dependence of the Dicke QPT, and show how to manipulate the
impurity-population in the cavity-BEC system. Finally, we shall
conclude our paper with discussions and remarks in the last section.

\section{\label{Sec:2} The impurity-doped Dicke model}

\begin{figure}[tbp]
\includegraphics[bb=15 416 360 770, width=3.2 in]{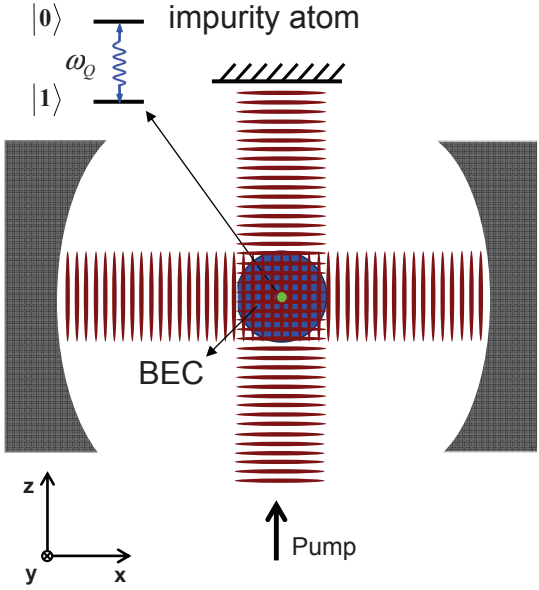}
\caption{(Color online) Schematic of the physical system under
consideration: An impurity qubit with energy separation $\omega_{Q}$
is doped into a atomic BEC in a ultrahigh-finesse cavity. Both the
impurity and BEC couple to a single cavity field and a transverse
pump field.} \label{fig1.eps}
\end{figure}

In this section, we establish the IDDM through combining cavity-BEC
and impurity-doped BEC techniques. Our proposed experimental setup
is indicated in Fig.~\ref{fig1.eps}. A two-level impurity atom
(qubit) with energy splitting $\omega_{Q}$ is doped in an atomic
BEC, which is confined in a ultrahigh-finesse optical cavity.

In the absence of the impurity atom, the cavity-BEC system under our
consideration is the same at that employed in the experiments to
observe the Dicke QPT \cite{Baumann1}. The cavity contains $N$
$^{87}\mathrm{Rb}$ condensed atoms interacting with a single cavity
model of frequency $\omega_{c}$ and a transverse pump field of
frequency $\omega_{p}$. The excited atoms may remit photons either
along or transverse to the cavity axis. This process couples the
zero momentum atomic ground state to the symmetric superposition
states of the $k$-momentum states. This yields an effective
two-level system. Suppose that the frequency $\omega_{c}$ and
$\omega_{p}$ are detuned far from the atomic resonance frequency
$\omega_{a}$, the excited atomic state can be adiabatically
eliminated. In this case, the single atom Hamiltonian of the system
under our consideration can be written as
\begin{eqnarray}
\hat{H}_{(1)} &=& \frac{\hat{p}_x^2 + \hat{p}_z^2}{2m}
+(U\cos^2k\hat{x}-\Delta_c)\hat{a}^\dag \hat{a}\nonumber\\
&&+V\cos^2(k\hat{z}) + \eta (\hat{a}^\dag +
\hat{a})\cos(k\hat{x})\cos(k\hat{z}).
\end{eqnarray}
Here the first term is the kinetic energy of the atom with momentum
operators $\hat{p}_{x,z}$. The second term describes the cavity
field, where $\hat{a}^{\dag}(\hat{a})$ is the creation
(annihilation) operator of the cavity field, which satisfy the
bosonic commutation $[\hat{a},\hat{a}^{\dag}]=1$, $U=
\frac{g_0^2}{\Delta_a}$ is the light shift induced by the atom where
$g_0$ is the atom-cavity coupling strength, $\Delta_a = \omega_{p}-
\omega_a$ and $\Delta_c=\omega_{p}-\omega_c$, $k$ is the
wave-vector, which is approximated to be equal on the cavity and
pump fields. The third term describe the potential along the
$z$-axis created by the pump field, the depth of the  potential $V =
\Omega_p^2/\Delta_a$ controlled by the maximum pump Rabi frequency
$\Omega_p$. The last term is the potential induced by the scattering
between the cavity field and the pump field, where $\eta = g_0
\Omega_p/\Delta_a$. The atom can be excited from the zero-momentum
state $|p_{x},p_{z}\rangle=|0,0\rangle$ to the $k$-momentum state
$|p_{x},p_{z}\rangle=\sum_{\upsilon_{1},\upsilon_{2}=\pm1}|\upsilon_{1}
k, \upsilon_{2} k\rangle$ through the scattering between the cavity
field and the pump field due to the conservation of momentum. So the
atomic field can be expanded in terms of two-mode approximation
$\hat{\Psi}=\Phi_{0}\hat{h}_{0}+\Phi_{1}\hat{h}_{1}$, where
$\hat{h}_{0}$ and $\hat{h}_{1}$ are bosonic operators and $\Phi_{0}$
($\Phi_{1}$) is the zero (k)-momentum single atom wave function.
Here $N=\hat{h}_{0}^{\dag}\hat{h}_{0}+\hat{h}_{1}^{\dag}\hat{h}_{1}$
represents the total number of condensed atoms, which holds
conservation in this paper. Substituting
$\hat{\Psi}=\Phi_{0}\hat{h}_{0}+\Phi_{1}\hat{h}_{1}$ into the second
quantization form
\begin{eqnarray}
\hat{H_{CB}} &= &\int\hat{\Psi}^\dag(x,z)\hat{H}_{(1)}\hat{\Psi}(x,z)\mathrm{d}x\,\mathrm{d}z\nonumber\\
&&+\frac{s}{2}\int\hat{\Psi}^\dag(x,z)\hat{\Psi}^\dag(x,z)\hat{\Psi}(x,z)\hat{\Psi}(x,z)\mathrm{d}x\,\mathrm{d}z,
\end{eqnarray}
where $s=4\pi a/m$, $a$ being s-wave scattering length. If one
introduces  the collective spin operators
$\hat{J_{z}}=(\hat{h}_{1}^{\dag}\hat{h}_{1}-\hat{h}_{0}^{\dag}\hat{h}_{0})/2$,
$\hat{J_{+}}=\hat{J_{-}}^{\dag}=\hat{h}_{1}^{\dag}\hat{h}_{0}$, up
to a constant term we obtain a extended Dicke model about the
cavity-BEC system
\begin{equation}
\label{1}\hat{H}_{CB}=\omega\hat{a}^{\dag}\hat{a}+\omega_{0}\hat{J}_{z}+\frac{\lambda
}{\sqrt{N}}\left(\hat{a}+\hat{a}^{\dag}\right)(\hat{J_{+}}+\hat{J_{-}})+\frac{\chi}{N}\hat{J_{z}}^{2},
\end{equation}
where the effective frequency of the cavity field
$\omega=-\Delta_{c}+NU_{0}/2$ and effective transition frequency
$\omega_{0}=2\omega_{r}+(N-1)(s_{1}-s_{0})/2$, where
$\omega_{r}=k^2/2m$ is recoil frequency and $s_{1(0)}=(4\pi
a_{1(0)}/m)\int d^{3}\vec{r}|\Phi_{1(0)}(\vec{r})|^{4}$ is the
interspecies coupling strength.
$\lambda=\sqrt{N}g_{0}\Omega_{p}/2\Delta$ is the coupling strength
induced by the cavity field and pump field, where $\Omega_{p}$
denotes the maximum pump Rabi frequency which can be adjusted by the
pump power. The nonlinear coupling strength is given by
$\chi=N[(s_{0}+s_{1})/2-s_{01}]$ with  $s_{01}=(4\pi a_{01}/m)\int
d^{3}\vec{r}|\Phi_{0}(\vec{r})|^{2}|\Phi_{1}(\vec{r})|^{2}$ being
interspecies coupling strength.

Next we consider interactions between the impurity qubit and the
cavity-BEC.  The impurity simultaneously interacts with the BEC ,
the cavity field, and the pump field. Firstly, we consider the
interaction between the impurity and the BEC. We assume that the
impurity interacts with the condensates via coherent collisions and
only the upper state $|0\rangle$ interacts with the condensate
considering its state-dependent trapped potential. Similar treatment
can also be found in the Ref. \cite {Ng2008}. Neglecting the
constant term, The impurity-BEC coupling Hamiltonian has the form
\begin{equation}
\hat{H}_{QB}=\kappa(\hat{\sigma}_{z}+1)\hat{J}_{z},
\end{equation}
where $\hat{\sigma}_{z}$ is the pauli operator of the impurity qubit
and $\kappa=\kappa_{0,0}-\kappa_{1,0}$, where $\kappa_{0(1),0}=(4\pi
a_{0(1),0}/m)\int
d^{3}\vec{r}|\Phi_{0(1)}(\vec{r})|^{2}|\varphi_{0}(\vec{r})|^{2}$ is
the coupling strength between the impurity and zero( $k$)- momentum
component BEC  with $\varphi_{0}(\vec{r})$ being the wave function
of the impurity in the upper state and $a_{0(1),0}$ being the $s-$
wave scattering length. Secondly, we consider the interactions
between the impurity qubit and the cavity field and the pump field.
The Hamiltonian of impurity qubit interacting with the cavity field
and the pump field reads as
\begin{equation}
\hat{H}_{QF}=\omega\hat{a}^{\dag}\hat{a}+\frac{\omega_{Q}}{2}\hat{\sigma}_{z}+g_{Q}\left(\hat{a}^{\dag}\hat{\sigma}_{-}
+\hat{a}\hat{\sigma}_{+}\right)+\Omega_{Q}(\hat{\sigma}_{+}+\hat{\sigma}_{-}),
\label{2}
\end{equation}
where $\sigma_{+} (\sigma_{-})$ is the raising  (lowering) operator
of the impurity qubit,  $g_{Q}$ the coupling strength between the
impurity qubit and the cavity field, $\Omega_{Q}$ the pump Rabi
frequency. Here we have made a rotating wave approximation. In the
far-detuning regime, one can use the Fr\"{o}hlich-Nakajima
transformation \cite{Fr1950,Naka1955} to make the Hamiltonian in Eq.
(\ref{2}) become the following expression
\begin{equation}
\label{3}\hat{H}_{QF}^{^{\prime
}}=\omega\hat{a}^{\dag}\hat{a}+\frac{\omega_{Q}^{^{\prime}}}{2}\hat{\sigma}_{z}
+\xi_{1}\hat{\sigma}_{z}\hat{a}^{\dag}\hat{a}+\xi_{2}\hat{\sigma}_{z}(\hat{a}^{\dag}+\hat{a}),
\end{equation}
where $\xi_{1}=g_{Q}^{2}/\Delta_{Q}$ and
$\xi_{2}=g_{Q}\Omega_{Q}/\Delta_{Q}$ with
$\Delta_{Q}=\omega_{Q}-\omega_{p}$.

Combining Eq.(3) with Eqs. (4) and (6), we arrive at the total
Hamiltonian of the IDDM
\begin{eqnarray}
\label{H1}\hat{H}&=&(\omega+\xi_{1}\hat{\sigma}_{z})\hat{a}^{\dag}\hat{a}
+[\omega_{0}+\kappa(\hat{\sigma}_{z}+1)]\hat{J}_{z}+\frac{\chi}{N}\hat{J_{z}}^{2}
+\frac{\omega_{Q}^{'}}{2}\hat{\sigma}_{z}\nonumber\\
&&+\frac{\lambda}{\sqrt{N}}\left(\hat{a}+\hat{a}^{\dag}\right)(\hat{J_{+}}+\hat{J_{-}})
+\xi_{2}\hat{\sigma}_{z}\left(\hat{a}+\hat{a}^{\dag}\right).
\end{eqnarray}
It is obvious that the Hamiltonian of the impurity-doped Dicke model
reduces to that of the original Dicke model when the
impurity-cavity-BEC interactions are switched off (i.e., $\kappa=0,
\xi_{1}=\xi_{2}=0$) and the atomic nonlinear interaction in the BEC
vanishes (i.e., $\chi=0$).

\section{\label{Sec:2} Impurity-induced Dicke quantum phase transition}

In this section, we study quantum phases and QPTs in the IDDM
proposed in the previous section. In order to understand QPTs, it is
necessary to investigate the ground-state properties for a many-body
system under consideration \cite{Sachdev1999}. For the
impurity-doped Dicke model in our proposal, its ground-state
properties can be analyzed in terms of Holstein-Primakoff
transformation \cite{Holstein1940} due to the large number of atoms
in the BEC. From the Hamiltonian (\ref{H1}), we can see that the
properties of the cavity-BEC system is related to the initial state
of the impurity qubit. We consider the impurity qubit as a control
tool over the cavity-BEC system which is the controlled target
system. In what follows, we will neglect  the nonlinear interaction
among condensed atoms to focus our attention on the influence of the
impurity on QPT.  Namely, we will take $\chi=0$ in the following
studies, which can be realized by Feshbach resonance techniques. Let
the impurity population $\delta=\langle\sigma_{z}\rangle$, and make
use of Holstein-Primakoff transformation to represent the angular
momentum operators as single-mode bosonic operators
($[\hat{c},\hat{c}^{\dag }]=1$)
\begin{eqnarray}
\hat{J}_{+}&=&\hat{c}^{\dag }\sqrt{N-\hat{c}^{\dag }\hat{c}},
\hspace{0.3cm}
\hat{J}_{-}=\sqrt{N-\hat{c}^{\dag }\hat{c}} \hat{c}, \nonumber\\
\hat{J}_{z}&=&\hat{c}^{\dag}\hat{c}-N/2,
\end{eqnarray}
after taking the mean value over a quantum state of the impurity
atom we can rewrite the Hamiltonian (\ref{H1}) as the following form
\begin{eqnarray}
\label{H2}\hat{H}^{'}&=&f_{1}\hat{a}^{\dag
}\hat{a}+f_{2}\hat{c}^{\dag }\hat{c}
+\xi_{2}\delta\left(\hat{a}+\hat{a}^{\dag}\right)\nonumber \\
&&+\frac{\lambda}{\sqrt{N}}\left(\hat{a}+\hat{a}^{\dag}\right)\left(\hat{c}^{\dag}\sqrt{N-\hat{c}^{\dag}\hat{c}}
+\sqrt{N-\hat{c}^{\dag}\hat{c}}\hat{c}\right),
\end{eqnarray}
where we have neglected a constant term, and effective frequencies
of the two coupled oscillator modes are given by
\begin{equation}
f_{1}=\omega +\xi _{1}\delta,\hspace{1cm}f_{2}=\omega _{0}+\kappa(1+
\delta).
\end{equation}

In order to describe the collective behaviors of the condensed atoms
and the photon, one can introduce new bosonic operators
$\hat{d}=\hat{a}+\sqrt{N}\alpha$ and $\hat{b}=\hat{c}-\sqrt{N}\beta$
\cite{Emary}, where $\alpha$ and $\beta$ are real numbers.
Substituting bosonic operators $\hat{d}$ and $\hat{b}$ into the
Hamiltonian (\ref{H2}) and neglecting terms with $N$ in the
denominator, the Hamiltonian (\ref{H2}) can be expanded by
\begin{eqnarray}
\label{H3} \hat{H}^{'}&=&NE_{0}+\sqrt{N}\hat{H}_{1}+\hat{H}_{2},
\end{eqnarray}
where we $E_{0}, \hat{H}_{1}$ and $\hat{H}_{2}$ are defined by
\begin{eqnarray}
E_{0} &=&f_{1}\alpha ^{2}+f_{2}\beta ^{2} -4\lambda K \alpha\beta, \\
\hat{H}_{1}&=&\left[ 2\lambda\alpha \left( K-\frac{\beta
^{2}}{K}\right) -f_{2}\beta \right](\hat{b}+\hat{b}^{\dag }) \nonumber\\
&&+\left( f_{1}\alpha-2\lambda K\beta\right)(\hat{d}+\hat{d}^{\dag})-2\xi_{2}\delta\alpha,\\
\hat{H}_{2} &=&f_{1}\hat{d}^{\dag }\hat{d} + \left(f_{2}
+\frac{2\lambda\alpha\beta}{K}\right)\hat{b}^{\dag}\hat{b}\nonumber\\
&&+\lambda \left( K-\frac{ \beta ^{2}}{K}\right)
(\hat{d}+\hat{d}^{\dag})(\hat{b}+\hat{b}^{\dag})\nonumber\\
&&+\frac{\beta^{3}}{2K^{3}}\left(\hat{b}+\hat{b}^{\dag}\right)^{2}+\xi_{2}\delta(\hat{d}+\hat{d}^{\dag}),
\end{eqnarray}
where we have introduced the parameter $K=\sqrt{1-\beta^{2}}$. The
collective excitation parameters $\alpha$ and $\beta$ can be
determined from the equilibrium conditions $\partial E_{0}/ \partial
\alpha=0$ and $\partial E_{0}/
\partial \beta=0$, which leads to the following two equations
\begin{equation}
f_{1}\alpha-2\lambda K\beta=0\label{p1}, \hspace{0.3cm}
2\lambda\alpha\left(K-\frac{\beta ^{2}}{K}\right)-f_{2}\beta =0,
\end{equation}
from which we can obtain an equation governing the fundamental
features of the QPT in the IDDM
\begin{equation}
\beta \left[8\lambda ^{2}\beta ^{2}+f_{1}f_{2}-4\lambda
^{2}\right]=0.
\end{equation}

Now we discuss quantum phases  and QPT in the impurity-doped Dicke
model.  When $f_{1}f_{2}\geq 4\lambda ^{2}$, from Eq. (16) we can
find $\alpha=\beta=0$ due to $\lambda ^{2}>0$. This means that both
the condensed atoms and the photon have not collective excitations.
Hence the cavity-BEC system is in the normal phase. However, when
$f_{1}f_{2}< 4\lambda ^{2}$, from Eqs. (15) and (16) we can obtain
the two nonzero collective excitation parameters
\begin{equation}
\label{ab} \alpha
^{2}=\frac{\lambda^{2}}{f_{1}^{2}}\left(1-\nu^{2}\right),\hspace{0.5cm}
\beta ^{2}=\frac{1}{2}\left(1-\nu \right),
\end{equation}
where we have let $\nu =f_{1}f_{2}/(4\lambda ^{2})$. Eq. (17)
implies that there exist macroscopic quantum population of the
collective excitations of the condensed atoms and the photon in the
IDDM. In this case, the cavity-BEC system is in the superradiant
phase. The Dicke QPT is  the QPT from the normal phase to the
superradiant phase.

From the QPT equation (16) we can see that there exist two
independent QPT parameters. One is the cavity-field-atom coupling
strength $\lambda$, another is the impurity population parameter
$\delta$.  This is one important difference between the IDDM and the
original Dicke model in which there is only one  QPT parameter, the
coupling strength  $\lambda$. Through the analysis below, we can see
that it is the introduction of the new QPT parameter  $\delta$ that
makes the IDDM to reveal new QPT characteristics which do not appear
in the original Dicke model. In the following, under the condition
$f_{1}\approx\omega$ due to $\omega\gg\xi _{1}\delta$, we
investigate the QPT in the IDDM for the three cases: (1) $\delta$ is
the QPT parameter with $\lambda$ being an arbitrary fixed parameter;
(2) $\lambda$ is the QPT parameter with $\delta$ being an arbitrary
fixed parameter;  (3) Both $\lambda$ and $\delta$ are independent
QPT parameters.

In the first case,  the impurity population $\delta$ is the QPT
parameter while the cavity-field-atom coupling strength $\lambda$ is
an arbitrary fixed parameter. So we can understand the QPT as the
impurity induced QPT. From the QPT equation (16) we can find that
the critical parameter $\delta_c$ at the QPT point satisfies the
following equation
\begin{equation}
\label{cri}
 \delta_{c}=\frac{4\lambda^{2}-\omega\omega_{0}}{\omega\kappa}-1,
\end{equation}
which indicates that there does always exist a critical impurity
population  $\delta_c$ for an arbitrary value of the
cavity-field-atom coupling strength $\lambda$.  From Eqs. (15) and
(16), we can find the two quantum phases of the normal phase and the
superradiant phase.  The normal phase is in the regime of $\delta
<\delta_c$  ($\delta >\delta_c$)  when $\kappa < 0$ ($\kappa >0$),
and we have $\alpha^2=\beta^2=0$. In the superradiant-phase regime,
we have nonzero collective excitations given by
\begin{eqnarray}
\alpha^{2}&=&\frac{\lambda^2}{\omega^{2}}-
\frac{(\omega_{0}+\kappa+\kappa\delta)^{2}}{16\lambda^2},\nonumber\\
\beta^{2}&=&\frac{1}{2}-\frac{\omega(\omega_{0}+\kappa+\kappa\delta
)}{4\lambda^2}.
\end{eqnarray}

It is interesting to note that in Eq. (18) the critical parameter
$\delta_c$ at the QPT point can vary continuously since the
cavity-field-atom coupling strength $\lambda$ can be manipulated
continuously.  This implies that the impurity-induced QPT in the
IDDM is a continuous QPT in which a quantum system can undergo a
continuous phase transition at the absolute zero of temperature as
some parameter entering its Hamiltonian is varied  continuously
\cite{son}. From the critical-point equation (18), we can see that
the impurity-induced Dicke QPT happens even in the weak coupling
regime of the cavity field and atoms. This is one of important
differences between the IDDM and the original Dicke model in which
the Dicke QPT appears only in the  strong coupling regime of the
cavity field and atoms.

We can determine the type of QPTs which happen in the impurity-doped
Dicke model through investigating the nonanalyticity of the scaled
energy  $E_0$  at the critical point in the thermodynamic limit
$N\longrightarrow \infty$. If the $n$th derivative shows nonanalytic
behavior then it is an $n$th order QPT. This has led researchers to
examine the behavior of different correlations near the critical
point, especially their analyticity properties as revealed by
differentiation. In Figure 2 we have plotted  the scaled
ground-state energy $E_{0}$ and its second derivative
$\partial^2E_{0}/\partial \delta^2$ with respect to  the QPT
parameter $\delta$. It is easy to know that the first derivative of
the scaled ground-state energy $E_{0}$ is continuous. From Fig. 2 we
can see that the second derivative  $\partial^2E_{0}/\partial
\delta^2$ is discontinuous at the quantum critical point
$\delta=\delta_c$. Therefore, we can conclude that the QPT induced
by the impurity is the second-order QPT.

In the second case, the cavity-field-atom coupling strength
$\lambda$ is the QPT parameter while the impurity population
$\delta$ is an arbitrary fixed parameter. So we can understand the
QPT as the cavity-field-atom coupling induced QPT. From the QPT
equation (16) we can find that the critical parameter $\lambda_c$ at
the QPT point satisfies the following equation
\begin{equation}
\label{con}
 4\lambda ^{2}_c-f_{1}f_{2}=0.
\end{equation}

Obviously, the cavity-field-atom coupling induced QPT  in the IDDM
is also a continuous QPT in which the QPT parameter can vary
continuously by changing the impurity population.  From equation
(20), we can find the critical coupling strength to be
\begin{equation}
\label{lac} \lambda_{c}= \frac{1}{2}\sqrt{\omega \omega _{0}+\omega
\kappa(1+ \delta)},
\end{equation}
which indicates that  the critical coupling strength $\lambda_{c}$
can continuously vary with the impurity population $\delta$ (
$-1\leq \delta \leq 1$). This means that the atom-field-coupling
induced QPT in the IDDM is a continuous QPT in which the critical
coupling strength $\lambda^{s}_{c}$  can take continuously an
infinite number of values when  the impurity population varies in
the regime $-1\leq \delta \leq 1$. Hence, there are an infinite
number of critical points of the QPT in the impurity-doped Dicke
model. This is an important difference between the impurity-doped
Dicke model and the original Dicke model in which there is only one
critical point of the QPT with the coupling strength being
$\lambda^{s}_{c}=\sqrt{\omega\omega_{0}}/2$, which can be recovered
from Eq. (21) when we take $\kappa=0$. From Eq. (\ref{lac}) we can
also see that  the attractive (repulsive) interaction $\kappa<0$
($\kappa>0$)  between the impurity and the condensed atoms can
decrease (increase) the critical coupling strength $\lambda_{c}$.
Therefore, we can realize the Dicke QPT in a broad range of the
coupling strength $\lambda$ through preparing various states of the
impurity atom. This provides a wide window to observe experimentally
the Dicke QPT.

\begin{figure}[tbp]
\label{GG}
 \includegraphics[clip=true,width=0.4\textwidth]{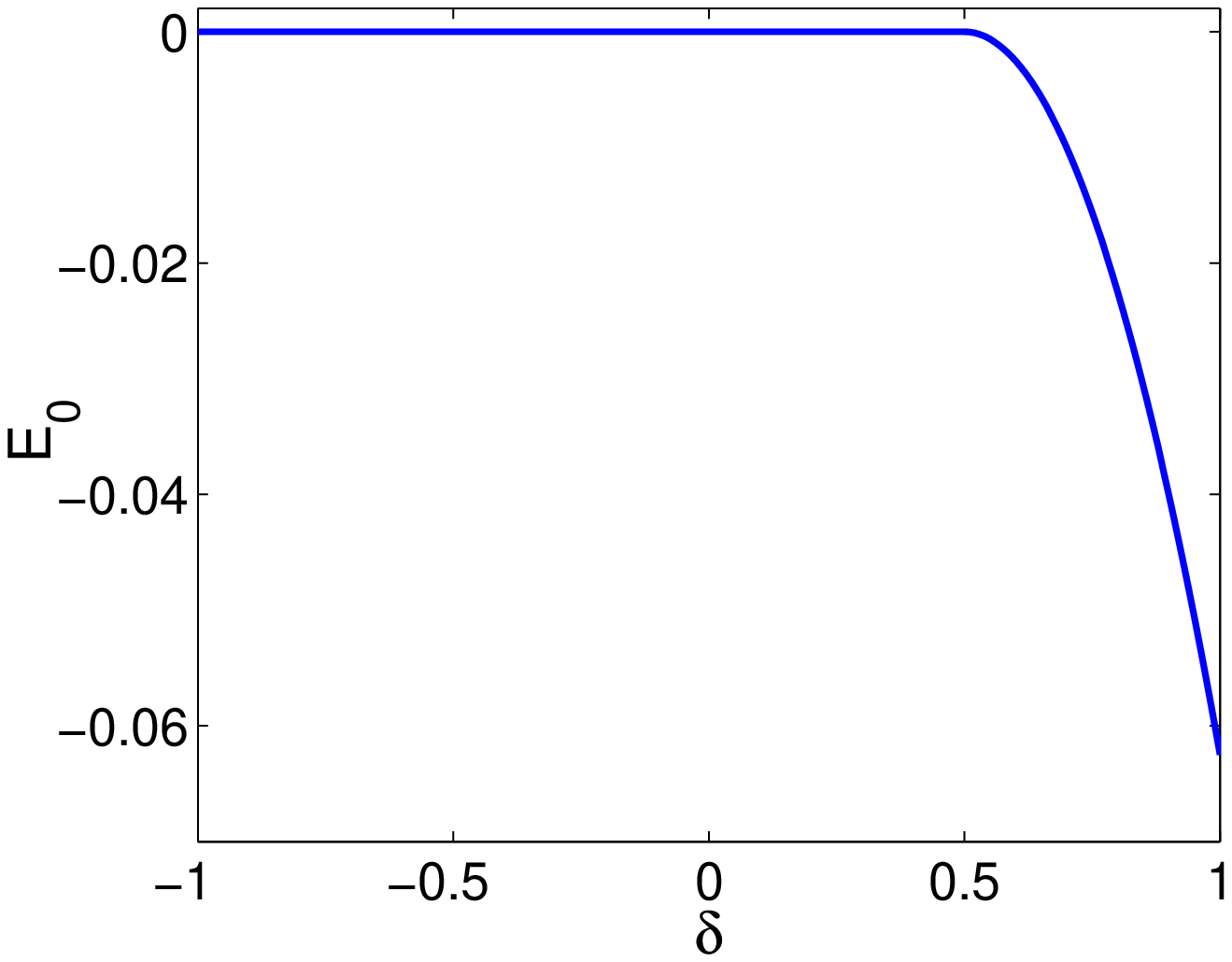}
    \includegraphics[clip=true,width=0.4\textwidth]{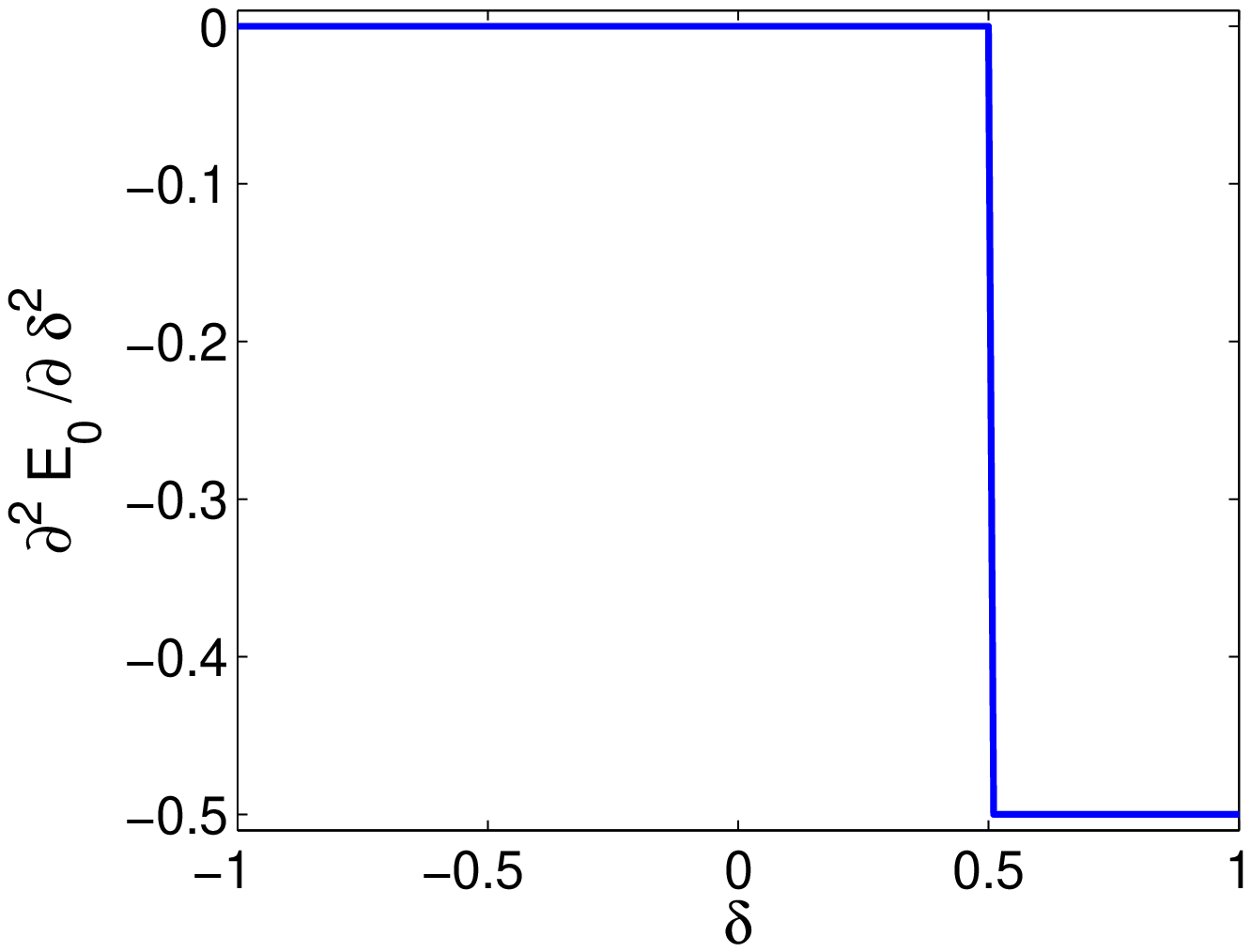}
  \caption{(Color online) The averaged ground-state energy  for the IDDM
and its second derivative in the infinite size limit. The related parameters are taken as
 $\omega=400$, $\kappa=-1/2$, and $\lambda=5$ in unit of $\omega_{0}$. The
discontinuity in the second-order derivative of the ground state
suggests a second-order quantum phase transition.}
\end{figure}

\begin{figure}[tbp]
\label{CP}
\includegraphics[clip=true,width=0.4\textwidth]{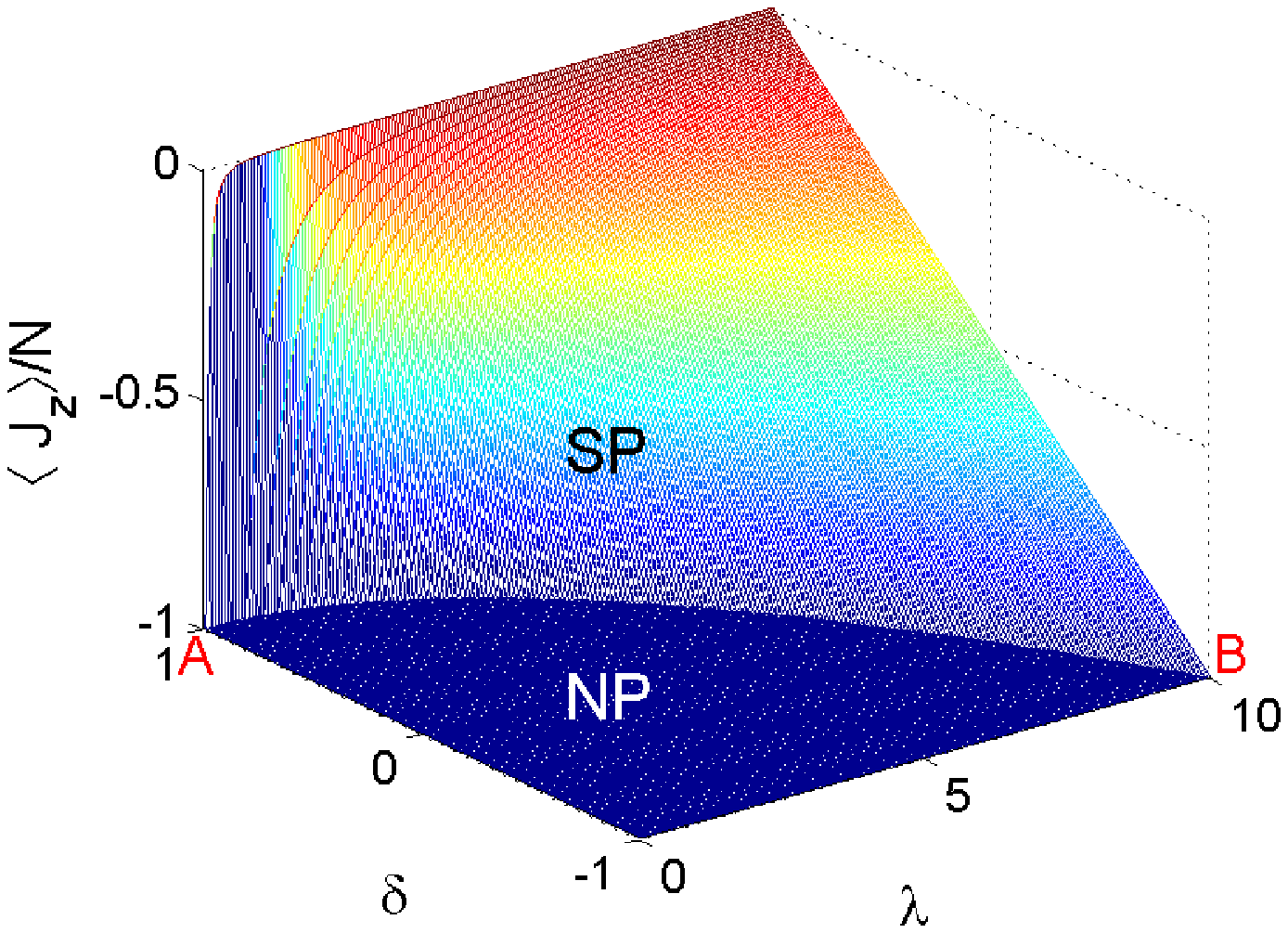}
\includegraphics[clip=true,width=0.4\textwidth]{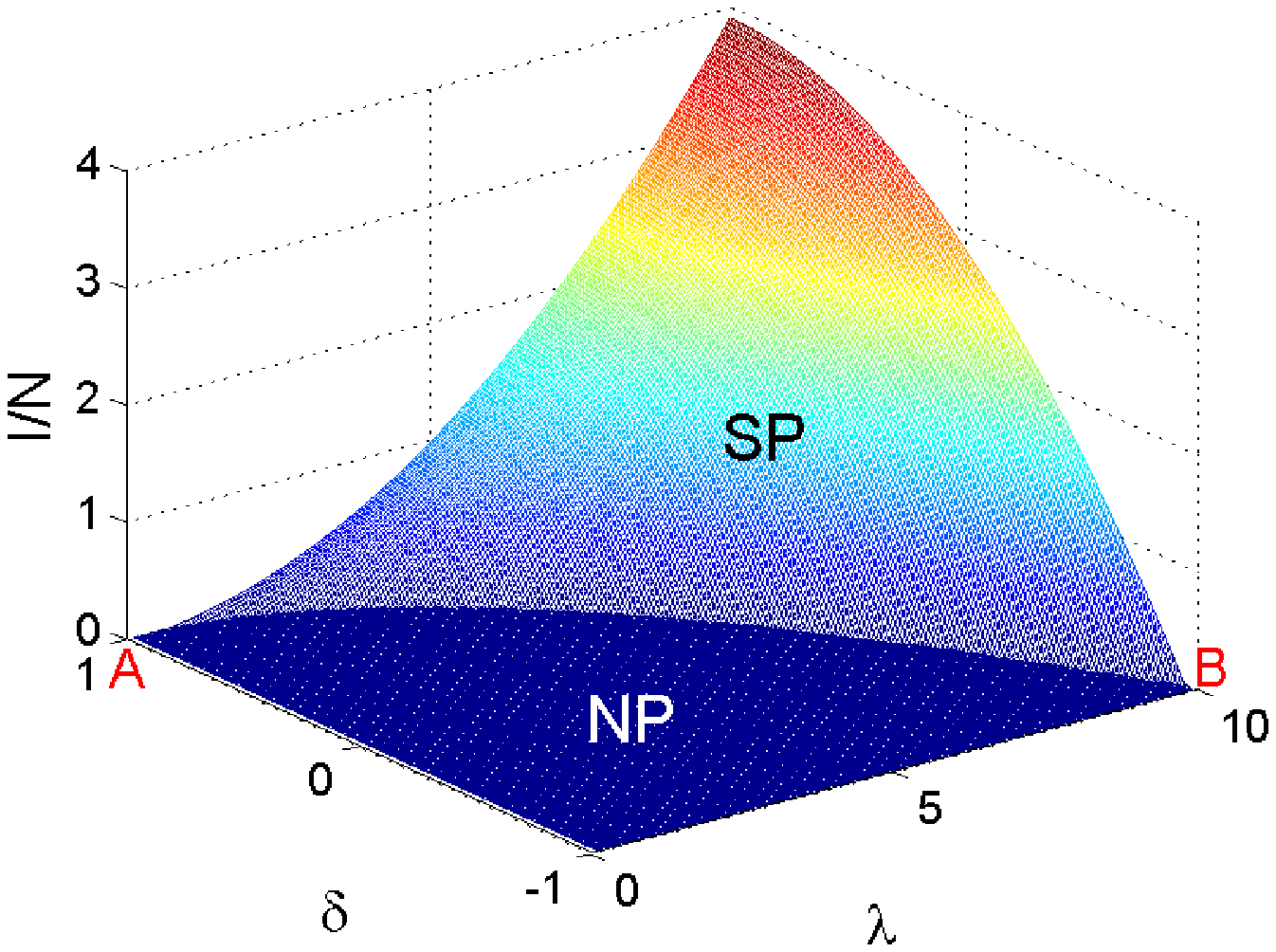}
\caption{\label{pi} (Color online) Phase diagrams described by  the
scaled population inversion of the BEC $\langle J_{z}\rangle/N$ and
the scaled intracavity-field intensity $I/N$ with respect to the
impurity population  $\delta$ and the coupling strength $\lambda$.
The related parameters are taken as $\omega=400$ and $\kappa=-1/2$
in unit of $\omega_{0}$.}
\end{figure}
 The third case is a general situation
in which two QPT parameters $\delta$ and $\delta$ vary
independently. In this case, nonzero collective excitations are
given by  Eq. (21). In the thermodynamic limit $N\longrightarrow
\infty$ we can obtain the scaled population inversion of BEC
$\langle J_{z}\rangle/N$ and the scaled intracavity intensity $I/N$
as
\begin{equation}
\langle J_{z}\rangle/N =\beta^{2}-1/2,\hspace{1cm}I/N=\alpha^2.
\end{equation}

We have plotted the phase diagrams of the IDDM for the general case
in Figure 3, which are described by the scaled population inversion
of BEC $\langle J_{z}\rangle/N$ and the scaled intracavity intensity
$I/N$ with respect to the impurity population $\delta$ and the
coupling strength $\lambda$. The related parameters are taken as
$\omega=400$ and $\kappa=-1/2$ in unit of $\omega_{0}$. From Figure
3 we can see that the normal phase is in the region of $\langle
J_{z}\rangle/N=0$ and $I/N=0$ while the superradiant phase is in the
nonzero region of $\langle J_{z}\rangle/$ and $I/N$. The Dicke QPT
happens at the critical curve $AB$ in the phase diagrams indicated
in Figure 3. The critical curve in the phase diagrams appears as the
intersection of the two phase regimes for the normal and
superradiant phases, and it can be described by the equation
\begin{equation}
\label{CC}
 \lambda^{2}+50\delta-50=0.
\end{equation}
The cavity-BEC is in normal-phase in the regime of $
\lambda^{2}+50\delta-50<0$ and in superradiant phase when
$\lambda^{2}+50\delta-50>0$. In superradiant phase, the collective
excitations increase with the QPT parameters $\delta$ and $\lambda$.

Finally, we show how to manipulate the impurity population, which is the key point to observe the Dicke QPT induced by the impurity atom.  In order to do this, We introduce  an auxiliary atom
outside the cavity, which is correlated with the impurity atom. We indicate that the impurity population can be controlled by making projective measurements upon the  auxiliary atom.
As an example, we consider the case of the impurity atom $A$ and the auxiliary atom
$B$  initially being in the well-known Werner state
\begin{equation}
\label{st} \rho=\frac{1-z}{4}\hat{I}+z|\Psi\rangle\langle\Psi|,
\hspace{0.2cm} 0\leq z \leq1,
\end{equation}
where $\hat{I}$ is the unit operator, $|\Psi\rangle$ is Bell state
$|\Psi\rangle=(|0\rangle_{A}|0\rangle_{B}+|1\rangle_{A}|1\rangle_{B})/\sqrt{2}$.
In this state, if one dose not measure the auxiliary atom, the
impurity population is zero, i.e.,
$\delta=\mathrm{Tr_{AB}}[\rho\hat{\sigma}^{A}_{z})]=0$.

We now introduce two orthogonal complete projection operators
\begin{equation}
\hat{\Pi}^B_{\pm}(\theta)=|\psi(\theta)\rangle^B_{\pm}\langle\psi(\theta)|,
\end{equation}
where $|\psi(\theta)\rangle_{\pm}$ are two  orthogonal
quantum states of the auxiliary atom
\begin{equation} |\psi(\theta)\rangle^B_{\pm}=\sin\theta|1\rangle \pm
\cos\theta|0\rangle
\end{equation}

For the initial state given by Eq. (24), after making the projective
measurements $\hat{\Pi}^B_{\pm}(\theta)$ upon the auxiliary atom
$B$, we can find that the impurity atom will collapse to the
following state
\begin{eqnarray}
\rho_{A}^{\pm}
=\frac{1-z}{2}\hat{I}+z|\psi(\theta)\rangle^{A}_{\pm}{\langle|\psi(\theta)}|.
\end{eqnarray}
 From Eq. (27) we can obtain
the impurity population
\begin{equation}
\label{cd} \delta_{\pm}=\pm z\cos2\theta.
\end{equation}

From Eq. (\ref{cd}), we can see that the impurity population depends
on the initially state parameter $z$ and the angle of the projection
measurement  $\theta$ upon the auxiliary atom.  Therefore, we can
manipulate the impurity population through making projective
measurements along different directions upon quantum states of the
auxiliary atom.   Eq. (\ref{cd}) indicates that the impurity
population vanishes when $z=0$, in which there does not exist
quantum correlation between the impurity atom and the auxiliary
atom. In this sense,   the nonzero impurity population is induced by
quantum correlation between the impurity atom and the auxiliary
atom.

\section{\label{Sec:4}Conclusions}

In conclusion, we have presented a generalized Dicke model, i.e.,
the IDDM, by the use of an impurity-doped cavity-Bose-Einstein
condensate. The original Dicke mode can be recovered under certain
conditions as a special case of the IDDM. We have shown that the
impurity atom can induce the Dicke QPT from the normal phase to
superradiant phase at a critic value of the impurity population. The
impurity-induced Dicke QPT  can be manipulated by the use of the
impurity population. We have proposed a scheme to control the
impurity population in the BEC through making quantum measurements
on an auxiliary atom outside the cavity, which is correlated to the
impurity atom in the BEC. We have found that the IDDM exhibits the
continuous Dicke QPT with an infinite number of critical points.
This multi-critical-point Dicke QPT is very different from the Dicke
QPT observed in the standard Dicke model with only one critical
point. In the IDDM, both the impurity atom and condensed atoms can
induce the Dicke QPT. It is the interaction between the
impurity-induced Dicke QPT and the cavity-field-BEC coupling induced
Dicke QPT that leads the appearance of multi-critical points in the
IDDM. These multi-critical points may be used as a resource for
processing quantum information \cite{Yuan,Zan}. We have predicted
that the impurity-induced Dicke QPT can happen in an arbitrary
coupling regime of the cavity field and atoms while the Dicke QPT in
the standard Dicke model occurs only in the strong coupling regime
of the cavity field and atoms. Hence, the IDDM reveals new regions
of the Dicke QPT. This opens a way to observe the Dicke QPT in the
intermediate and even weak coupling regime of the cavity field and
atoms.  Based on current experimental developments, we believe that
it is possible to observe experimentally the impurity-induced Dicke
QPT  by measuring the atomic population or the mean photon number of
the cavity field. The experimental realization of the scheme
proposed in the present paper deserves further investigation.

\acknowledgments This work was supported by the National Fundamental
Research Program of China under Grant No. 2007CB925204, the National
Natural Science Foundation of China under Grant No. 10775048 and No.
11375060, the Program for Changjiang Scholars and Innovative
Research Team in University under Grant No. IRT0964, the Research
Fund of Hunan Provincial Education Department Grant No. 08W012, and
Hunan Provincial Innovation Foundation for Postgraduate under Grant
No. CX2013B220.

\end{document}